\documentclass[11pt]{revtex4}
\usepackage{mathtools}
\usepackage{hyperref}
\usepackage{pdflscape}
\usepackage{graphics}
\usepackage{subfigure}
\usepackage{epsfig}
\usepackage{amsmath,amsfonts}
\usepackage{amssymb}
\usepackage{color}
\usepackage{ulem}
\usepackage{amsmath}
\allowdisplaybreaks

\begin{document}
\title{Polarization modes of gravitational waves in general symmetric teleparallel gravity}

\author{Yu-Qi Dong$^{a,b}$}
\email{dongyq2023@lzu.edu.cn}

\author{Xiao-Bin Lai$^{a,b}$}
\email{laixb2024@lzu.edu.cn}

\author{Yu-Zhi Fan$^{a,b}$}
\email{fanyzh2024@lzu.edu.cn}

\author{Yu-Xiao Liu$^{a,b}$}
\email{liuyx@lzu.edu.cn (corresponding author)}

\affiliation{$^{a}$ Lanzhou Center for Theoretical Physics, Key Laboratory of Theoretical Physics  of Gansu Province, Key Laboratory of Quantum Theory and Applications of MoE, Gansu Provincial Research Center for Basic Disciplines of Quantum Physics, Lanzhou University, Lanzhou 730000, China\\
	$^{b}$ Institute of Theoretical Physics \& Research Center of Gravitation, Lanzhou University, Lanzhou 730000, China}

\begin{abstract}
\textbf{Abstract:} In this paper, we investigate the polarization modes of gravitational waves within the most general symmetric teleparallel gravity theory that allows for second-order field equations We consider both scenarios where test particles either carry or do not carry a hypermomentum charge. Our findings reveal the existence of tensor, vector, and scalar modes of gravitational waves. Firstly, the theory supports the + and $\times$ tensor modes propagating at the speed of light. Secondly, in the case where particles do not carry hypermomentum, vector modes propagating at the speed of light exist only within a very specific parameter space. However, when particles do carry hypermomentum, there are two shear modes that propagate at the speed of light, while the vector-$x$ and vector-$y$ modes emerge only under very specific conditions. Thirdly, in the presence of hypermomentum, there is always a longitudinal mode propagating at the speed of light. The universal existence of the shear modes and the longitudinal mode in the presence of hypermomentum is a key feature of symmetric teleparallel gravity, distinguishing it from the Riemannian framework through gravitational wave polarization detection. We also analyze the polarization modes in two widely studied special theories: $f(Q)$ theory and quadratic non-metricity theory. Our study reveals that, within the $f(Q)$ gravity framework, it is crucial to assume that matter fields are independent of the connection, as any dependence would lead to unphysical results.

\end{abstract}
	
\maketitle

\section{Introduction}
\label{sec: intro} 

The recently proposed geometrical trinity of general relativity \cite{Lavinia Heisenberg1, Lavinia Heisenberg2,Salvatore Capozziello Vittorio De Falco Carmen Ferrara,Salvatore Capozziello Vittorio De Falco Carmen Ferrara2} illustrates that, at the level of classical field equations, gravitational  theories equivalent to general relativity can be formulated within two geometric frameworks that differ fundamentally from Riemannian geometry. One such framework is metric teleparallel geometry \cite{Sebastian Bahamonde}, which departs from Riemannian geometry by requiring that the spacetime curvature $R^{\mu}_{~\nu\lambda\rho}$ and the non-metricity $Q_{\lambda\mu\nu}$ vanish, while gravity arises purely from the torsion $T^{\lambda}_{~\mu\nu}$. The other framework, known as symmetric teleparallel geometry \cite{Lavinia Heisenberg3}, attributes gravity entirely to the non-metricity $Q_{\lambda\mu\nu}$, while both the curvature $R^{\mu}_{~\nu\lambda\rho}$ and the torsion $T^{\lambda}_{~\mu\nu}$ are set to zero.

Since the field equations of these theories remain equivalent, there is no fundamental reason to favor Riemannian geometry over these alternative geometric descriptions of gravity. It is entirely conceivable that in an alternate parallel universe, an ``Einstein" might have formulated a ``general relativity" based on teleparallel gravity instead. Despite their formal equivalence, the distinct geometric foundations of these theories lead to different physical interpretations, offering deeper insights into the nature of gravity. Furthermore, their differences at the quantum level may provide new perspectives for tackling challenging issues in quantum gravity that remain unresolved within general relativity \cite{Lavinia Heisenberg4,Ruben Aldrovandi,Lavinia Heisenberg5}. Finally, considering the modifications of gravity within these geometric frameworks can effectively bypass the constraints of Lovelock's theorem \cite{David Lovelock1,David Lovelock2}, leading to new classes of modified gravitational theories and providing novel approaches to addressing problems in quantum gravity and cosmology \cite{Sebastian Bahamonde,Lavinia Heisenberg3}. For the above reasons, the geometrical trinity of general relativity and its modifications have received widespread attention and research.

In this paper, we focus on various gravitational theories within the symmetric teleparallel geometry framework and generally analyze the polarization modes of gravitational waves. The era of gravitational wave astronomy has arrived \cite{Abbott1,Abbott2,Abbott3,Abbott4,Abbott5}, which means that we can now and in the near future test various candidate modified gravitational theories through high-precision gravitational wave detection \cite{J. Aasi,F. Acernese,T. Akutsu,M. Punturo,D. Reitze,P. Amaro-Seoane,Z. Luo,Jun Luo,Lijing Shao,H. B. Jin,Y. Y. Wang}. Polarization, as a fundamental property of gravitational waves, is naturally one of the key focuses of gravitational wave detection. In has been proven in Ref. \cite{Eardley} that there are at most six independent gravitational wave polarization modes in a four-dimensional  gravitational theory with matter fields coupling only to the metric. Furthermore, our previous research indicates that in a torsionless spacetime, test particles carrying hypermomentum charge may detect two additional polarization modes beyond the six known ones \cite{Yu-Qi Dong}. Different theories predict different polarization modes of gravitational waves, so the detection of gravitational wave polarizations can be used to test various gravitational theories. This highlights the necessity of the research topic of this paper. There have already been extensive theoretical and technical studies of gravitational wave polarization modes \cite{Yu-Qi Dong,f(R,fT,Orfeu Bertolami,Horndeski0,TeVeS,dCS and EdGB,Y.Dong,Y.Dong3,Y.Dong4,Y.Dong5,Y. Fan,S. Bahamonde,L.Shao,Shaoqi Hou,Xiao-Bin Lai,Miguel Barroso Varela,S Capozziello,Kristen Schumacher1,Kristen Schumacher2,Ismail Soudi,Manuel Hohmann,Ligong Bian}.

Considering that in symmetric teleparallel gravity, the connection is not a priori expressed as a function of the metric, the existence of hypermomentum is naturally possible in the theory. Based on this, we consider two cases: one without hypermomentum and one with hypermomentum. In both cases, we analyze the polarization modes of gravitational waves in the most general symmetric teleparallel gravity that leads to second-order field equations. The analytical approach in this paper is based on the general framework developed in our previous research \cite{Y.Dong4,Y.Dong5,Yu-Qi Dong}. 

This paper is organized as follows: In Sec. \ref{sec: 2}, we discuss the linearized field equations of symmetric teleparallel gravity, which serve as the fundamental equations required for solving gravitational wave polarization modes. In Sec. \ref{sec: 3}, we introduce the relevant concepts of gravitational wave polarization modes and explain their connection with gauge-invariant perturbations. These gauge-invariant perturbations significantly simplify the linearized field equations. In Sec. \ref{sec: 4}, we analyze the polarization modes of gravitational waves in general symmetric teleparallel gravity. In Sec. \ref{sec: 5}, we specifically provide concrete examples of gravitational wave polarization modes in $f(Q)$ theory and quadratic non-metricity theory. These two theories have attracted significant attention and are representative. Finally, Sec. \ref{sec: 6} presents the conclusion.

We adopt natural units with $c=G=1$ and use the metric signature $(-,+,+,+)$. Greek indices $(\mu, \nu, \lambda, \rho)$  span four-dimensional spacetime $(0, 1, 2, 3)$, while Latin indices $(i, j, k, l)$ cover three-dimensional spatial directions $(+x, +y, +z)$, ranging from $(1, 2, 3)$. The symbol $\mathring{\Gamma}^{\lambda}_{~\mu\nu}$ represents a general connection, whereas $\Gamma^{\lambda}_{~\mu\nu}$ denotes the torsionless connection and $\widehat{\Gamma}^{\lambda}_{~\mu\nu}$ denotes the Levi-Civita connection. For any quantity 
$A$ dependent on the connection, we define $\widehat{A}$
as its value when the connection is replaced by $\widehat{\Gamma}^{\lambda}_{~\mu\nu}$. In the linearized theory, indices are raised and lowered by the Minkowski metric $\eta_{\mu\nu}$.

\section{Linearized symmetric teleparallel gravity}
\label{sec: 2}

Gravitational waves are extremely weak. Therefore, the study of gravitational wave polarization modes only requires knowing their linear perturbation equations in a given flat spacetime background. For this reason, this section discusses the linearized symmetric teleparallel gravity theory.

In general metric-affine geometry, there exist three geometrically significant quantities related to parallel transport: the curvature $R^{\mu}_{~\nu\rho\sigma}$, the torsion $T^{\lambda}_{~\mu\nu}$, and the non-metricity $Q_{\lambda\mu\nu}$. They are defined as: 
\begin{eqnarray}
	\label{RTQ}
    R^{\mu}_{~\nu\rho\sigma} &\coloneqq& \partial_{\rho}\mathring{\Gamma}^{\mu}_{~\nu\sigma}-\partial_{\sigma}\mathring{\Gamma}^{\mu}_{~\nu\rho}+\mathring{\Gamma}^{\mu}_{~\tau\rho}\mathring{\Gamma}^{\tau}_{~\nu\sigma}-\mathring{\Gamma}^{\mu}_{~\tau\sigma}\mathring{\Gamma}^{\tau}_{~\nu\rho},
    \nonumber \\
    T^{\lambda}_{~\mu\nu} &\coloneqq& \mathring{\Gamma}^{\lambda}_{~\mu\nu}-\mathring{\Gamma}^{\lambda}_{~\nu\mu},
    \quad
    Q_{\lambda\mu\nu} \coloneqq  \nabla_{\lambda}g_{\mu\nu}.
\end{eqnarray}	
Symmetric teleparallel gravity a priori requires $R^{\mu}_{~\nu\rho\sigma}=T^{\lambda}_{~\mu\nu}=0$. Therefore, the action of general symmetric teleparallel gravity can be expressed by introducing Lagrange multipliers \cite{Sebastian Bahamonde,Lavinia Heisenberg3,Jose Beltran Jimenez}:
\begin{eqnarray}
	\label{general action}
	S=S_{g}\left[g_{\mu\nu},\mathring{\Gamma}^{\lambda}_{~\mu\nu}\right]
	+S_{m}\left[g_{\mu\nu},\mathring{\Gamma}^{\lambda}_{~\mu\nu},\Psi_{a}\right]
	+\int d^{4}x \sqrt{-g} ~
	\left(
	\mathfrak{p}_{\mu}^{~\nu\lambda\rho}R^{\mu}_{~\nu\lambda\rho}
	+\mathfrak{q}_{\lambda}^{~\mu\nu}T^{\lambda}_{~\mu\nu}
	\right).
\end{eqnarray}	
Here, $S_{g}$ is the action describing the gravitational part, while $S_{m}$ is the action describing the matter field $\Psi_{a}$. In the geometric framework of symmetric teleparallel gravity, there is no reason to require $S_{m}$ to be independent of the connection, thus naturally necessitating the consideration of the possible existence of hypermomentum. $\mathfrak{p}_{\mu}^{~\nu\lambda\rho}$ and $\mathfrak{q}_{\lambda}^{~\mu\nu}$ are Lagrange multipliers, which are tensors. Considering the symmetry of the indices in $R^{\mu}_{~\nu\lambda\rho}$ and $T^{\lambda}_{~\mu\nu}$, $\mathfrak{p}_{\mu}^{~\nu\lambda\rho}$ is antisymmetric with respect to the indices $(\lambda, \rho)$, while $\mathfrak{q}_{\lambda}^{~\mu\nu}$ is antisymmetric with respect to the indices $(\mu, \nu)$.
As long as we note that the equation obtained by varying the action (\ref{general action}) with respect to $\mathfrak{q}_{\lambda}^{~\mu\nu}$ only requires the torsion $T^{\lambda}_{~\mu\nu}=0$, and that the antisymmetric part of the indices $(\mu, \nu)$ in the equation obtained by varying the action (\ref{general action}) with respect to the connection $\mathring{\Gamma}^{\lambda}_{~\mu\nu}$ only determines the value of the Lagrange multiplier $\mathfrak{q}_{\lambda}^{~\mu\nu}$, it can be proven that the action (\ref{general action}) is equivalent to the following action:
\begin{eqnarray}
	\label{general action 2}
	S=S_{g}\left[g_{\mu\nu},{\Gamma}^{\lambda}_{~\mu\nu}\right]
	+S_{m}\left[g_{\mu\nu},{\Gamma}^{\lambda}_{~\mu\nu},\Psi_{a}\right]
	+\int d^{4}x \sqrt{-g} ~
	\mathfrak{p}_{\mu}^{~\nu\lambda\rho}R^{\mu}_{~\nu\lambda\rho},
\end{eqnarray}	
where the general connection in all cases is replaced by the torsionless connection.

A good physical theory allows us to reasonably neglect the gravitational backreaction of test particles used for detecting gravitational waves on the spacetime background. Therefore, when studying the polarizations of gravitational waves, the vacuum case without matter can always be considered, and the spacetime background can be set to flat. We assume that the flat spacetime background is 
\begin{eqnarray}
	\label{flat spacetime background}
	g_{\mu\nu}=\eta_{\mu\nu},\quad
	\Gamma^{\lambda}_{~\mu\nu}=0,\quad
	\mathfrak{p}_{\mu}^{~\nu\lambda\rho}=\bar{\mathfrak{p}}_{\mu}^{~\nu\lambda\rho}.
\end{eqnarray}
The background equations obtained from the variation of the action with respect to the connection can be shown to satisfy 
\begin{eqnarray}
	\label{background equation}
	\partial_{\rho}\bar{\mathfrak{p}}_{\lambda}^{~(\mu\nu)\rho}=0,
\end{eqnarray}
as long as it is noted that $\eta_{\mu\nu}$ has only two indices and $\partial_{\lambda}\eta_{\mu\nu}=0$. Then, we can take the background perturbations
\begin{eqnarray}
	\label{background perturbations}
	g_{\mu\nu}=\eta_{\mu\nu}+h_{\mu\nu},\quad
	\Gamma^{\lambda}_{~\mu\nu}=-\Sigma^{\lambda}_{~\mu\nu},\quad
	\mathfrak{p}_{\mu}^{~\nu\lambda\rho}=\bar{\mathfrak{p}}_{\mu}^{~\nu\lambda\rho}+p_{\mu}^{~\nu\lambda\rho}.
\end{eqnarray}
To obtain the linear perturbation equations, it is sufficient to consider the second-order terms in the perturbative expansion of the action (\ref{general action 2}). By varying the second-order perturbation action with respect to the perturbations, we obtain the linear perturbation equations:
\begin{eqnarray}
	\label{linear perturbation equations metric}
	\mathcal{F}_{\mu\nu}&=&0,
	\\
	\label{linear perturbation equations connection}
	\mathcal{G}_{\lambda}^{~\mu\nu}&=&
	-2\bar{\mathfrak{p}}_{\lambda}^{~\alpha\beta(\nu}\Sigma^{\mu)}_{~\alpha\beta}
	-2\bar{\mathfrak{p}}_{\alpha}^{~(\mu\nu)\beta}\Sigma^{\alpha}_{~\lambda\beta}
	-\partial_{\rho}p_{\lambda}^{~(\mu\nu)\rho}-\partial_{\rho}\left(h~ \bar{\mathfrak{p}}_{\lambda}^{~(\mu\nu)\rho}\right),
	\\
	\label{linear perturbation equations R}
	\partial_{\rho}\Sigma^{\mu}_{~\nu\lambda}&=&\partial_{\lambda}\Sigma^{\mu}_{~\nu\rho}.
\end{eqnarray}
Here, $h\coloneqq \eta^{\mu\nu}h_{\mu\nu}$, the expression for $\mathcal{F}_{\mu\nu}$ and $\mathcal{G}_{\lambda}^{~\mu\nu}$ are obtained by varying the second-order perturbation action of $S_{g}$ with respect to $h^{\mu\nu}$ and $\Sigma^{\lambda}_{~\mu\nu}$, respectively. Equation (\ref{linear perturbation equations R}) corresponds to the linear term of $R^{\mu}_{~\nu\lambda\rho}=0$. The above linearized field equations are actually equivalent to the following equations:
\begin{eqnarray}
	\label{linear perturbation equations metric new}
	\mathcal{F}_{\mu\nu}&=&0,
	\\
	\label{linear perturbation equations connection new}
	\partial_{\mu}\partial_{\nu}\mathcal{G}_{\lambda}^{~\mu\nu}&=&0,
	\\
	\label{linear perturbation equations R new}
	\partial_{\rho}\Sigma^{\mu}_{~\nu\lambda}&=&\partial_{\lambda}\Sigma^{\mu}_{~\nu\rho}.
\end{eqnarray}
This is because, by using Eqs. (\ref{background equation}), (\ref{linear perturbation equations connection}), (\ref{linear perturbation equations R}) and the symmetry of the indices, we can derive Eq. (\ref{linear perturbation equations connection new}). In addition, for every solution of 
$h_{\mu\nu}$ and $\Sigma^{\lambda}_{~\mu\nu}$ that satisfies Eqs. (\ref{linear perturbation equations metric new})-(\ref{linear perturbation equations R new}), there always exists $p_{\lambda}^{~\mu\nu\rho}$ (generally not unique, but this is not a problem) such that $h_{\mu\nu}, \Sigma^{\lambda}_{~\mu\nu}$, and $p_{\lambda}^{~\mu\nu\rho}$ satisfy Eqs. (\ref{linear perturbation equations metric})-(\ref{linear perturbation equations R}). Equations (\ref{linear perturbation equations metric new})-(\ref{linear perturbation equations R new}) are the equations we need.

In order to study the polarization modes of gravitational waves in the most general symmetric teleparallel gravity theory that can derive second-order field equations, we need to provide the most general second-order perturbation action $S_{g}^{(2)}$ for $S_{g}$. This has already been presented in our previous paper \cite{Yu-Qi Dong}, and it is
\begin{eqnarray}
	\label{the most general second-order perturbation action}
	S_{g}^{(2)}=S^{(2)}_{0}+S^{(2)}_{1}+S^{(2)}_{2}=\int d^4 x  \left(\mathcal{L}_{0}+\mathcal{L}_{1}+\mathcal{L}_{2}\right),
\end{eqnarray}
where
\begin{eqnarray}
	\label{L0}
	\mathcal{L}_{0}
	\!&=&\!
	\left(2B_{(1)}+E_{(1)}\right)\eta^{\alpha\beta}\Sigma^{\lambda}_{~\lambda\mu}\Sigma^{\mu}_{~\alpha\beta}
	+\left(B_{(1)}+D_{(1)}+\frac{1}{2}E_{(1)}\right)\eta^{\mu\nu}\Sigma^{\lambda}_{~\lambda\mu}\Sigma^{\rho}_{~\rho\nu}
	\nonumber \\
	\!&+&\!
	\frac{1}{2}C_{(1)}\eta_{\lambda\rho}\eta^{\mu\alpha}\eta^{\nu\beta}\Sigma^{\lambda}_{~\mu\nu}\Sigma^{\rho}_{~\alpha\beta}
	+\frac{1}{2}E_{(1)}\eta_{\lambda\rho}\eta^{\mu\nu}\eta^{\alpha\beta}\Sigma^{\lambda}_{~\mu\nu}\Sigma^{\rho}_{~\alpha\beta}
	\nonumber \\
	\!&+&\!
	\left(A_{(1)}+\frac{1}{2}C_{(1)}\right)\eta^{\nu\rho}\Sigma^{\lambda}_{~\mu\nu}\Sigma^{\mu}_{~\lambda\rho},
\end{eqnarray}
\begin{eqnarray}
	\label{L1}
	\mathcal{L}_{1}
	\!&=&\!
	A_{(1)}\eta^{\mu\alpha}\eta^{\nu\beta}\left(\partial_{\lambda}h_{\mu\nu}\right)\Sigma^{\lambda}_{~\alpha\beta}
	+B_{(1)}\eta^{\alpha\beta}\left(\partial_{\lambda}h\right)\Sigma^{\lambda}_{~\alpha\beta}
	+C_{(1)}\eta^{\nu\alpha}\left(\partial^{\beta}h_{\mu\nu}\right)\Sigma^{\mu}_{~\alpha\beta}
	\nonumber\\
	\!&+&\!
	D_{(1)}\eta^{\lambda\rho}\left(\partial_{\lambda}h\right)\Sigma^{\mu}_{~\mu\rho}
	+E_{(1)}\eta^{\lambda\nu}\eta^{\alpha\beta}\left(\partial_{\lambda}h_{\mu\nu}\right)\Sigma^{\mu}_{~\alpha\beta}
	\nonumber \\
	\!&+&\!
	\left(2B_{(1)}+E_{(1)}\right)\eta^{\mu\rho}\eta^{\lambda\nu}\left(\partial_{\lambda}h_{\mu\nu}\right)\Sigma^{\sigma}_{~\rho\sigma},
\end{eqnarray}
\begin{eqnarray}
	\label{L2}
	\mathcal{L}_{2}
	\!&=&\!
	A_{(2)}\left(\Box h_{\mu\nu}\right)h^{\mu\nu}
	+
	\left(-A_{(2)}-\frac{1}{4}C_{(1)}+\frac{1}{4}A_{(1)}+\frac{1}{4}B_{(1)}-\frac{1}{4}D_{(1)}\right)\left(\Box h\right)h
	\nonumber \\
	\!&+&\!
	\left(2A_{(2)}-\frac{1}{2}A_{(1)}+\frac{1}{2}C_{(1)}-B_{(1)}\right)\left(\partial_{\mu}\partial_{\nu}h\right)h^{\mu\nu}
	\nonumber \\
	\!&+&\!
	\left(-2A_{(2)}-\frac{1}{2}C_{(1)}-\frac{1}{2}E_{(1)}\right)\left(\partial_{\mu}\partial_{\nu}h^{\mu\rho}\right)h^{\nu}_{~\rho}
	\nonumber \\
	\!&+&\!
	B_{(2)}\eta^{\alpha\beta}\left(\Box \Sigma^{\lambda}_{~\lambda\mu}\right)\Sigma^{\mu}_{~\alpha\beta}
	+C_{(2)}\eta^{\mu\nu}\left(\Box\Sigma^{\lambda}_{~\lambda\mu}\right)\Sigma^{\rho}_{~\rho\nu}
	+D_{(2)}\eta_{\lambda\rho}\eta^{\mu\alpha}\eta^{\nu\beta}\left(\Box\Sigma^{\lambda}_{~\mu\nu}\right)\Sigma^{\rho}_{~\alpha\beta}
	\nonumber \\
	\!&+&\!
	E_{(2)}\eta_{\lambda\rho}\eta^{\mu\nu}\eta^{\alpha\beta}\left(\Box\Sigma^{\lambda}_{~\mu\nu}\right)\Sigma^{\rho}_{~\alpha\beta}
	+F_{(2)}\eta^{\nu\rho}\left(\Box\Sigma^{\lambda}_{~\mu\nu}\right)\Sigma^{\mu}_{~\lambda\rho}
	\nonumber \\
	\!&-&\!
	2E_{(2)}\eta^{\alpha\beta}\eta_{\lambda\rho}\left(\partial^{\mu}\partial^{\nu}\Sigma^{\lambda}_{~\mu\nu}\right)\Sigma^{\rho}_{~\alpha\beta}
	-B_{(2)}\left(\partial^{\mu}\partial^{\nu}\Sigma^{\lambda}_{~\mu\nu}\right)\Sigma^{\rho}_{~\lambda\rho}
	\nonumber \\
	\!&+&\!
	H_{(2)}\eta^{\lambda\rho}\left(\partial_{\mu}\partial^{\nu}\Sigma^{\mu}_{~\nu\lambda}\right)\Sigma^{\alpha}_{~\rho\alpha}
	+I_{(2)}\eta^{\alpha\beta}\left(\partial_{\mu}\partial^{\nu}\Sigma^{\mu}_{~\nu\lambda}\right)\Sigma^{\lambda}_{~\alpha\beta}
	+J_{(2)}\eta^{\lambda\rho}\eta^{\alpha\beta}\left(\partial_{\mu}\partial_{\nu}\Sigma^{\mu}_{~\lambda\rho}\right)\Sigma^{\nu}_{~\alpha\beta}
	\nonumber \\
	\!&+&\!
	\left(\frac{1}{2}B_{(2)}-\frac{1}{2}H_{(2)}+F_{(2)}+\frac{1}{2}I_{(2)}+K_{(2)}\right)\eta^{\lambda\alpha}\eta^{\rho\beta}\left(\partial_{\mu}\partial_{\nu}\Sigma^{\mu}_{~\lambda\rho}\right)\Sigma^{\nu}_{~\alpha\beta}
	\nonumber \\
	\!&+&\!
	\left(-D_{(2)}+E_{(2)}\right)\eta_{\lambda\rho}\eta^{\alpha\beta}\left(\partial^{\mu}\partial^{\nu}\Sigma^{\lambda}_{~\mu\alpha}\right)\Sigma^{\rho}_{~\nu\beta}
	+K_{(2)}\left(\partial^{\mu}\partial^{\nu}\Sigma^{\lambda}_{~\mu\rho}\right)\Sigma^{\rho}_{~\nu\lambda}
	\nonumber \\
	\!&+&\!
	\left(\frac{1}{2}B_{(2)}+\frac{1}{2}I_{(2)}+J_{(2)}-\frac{1}{2}H_{(2)}-C_{(2)}\right)\left(\partial^{\mu}\partial^{\nu}\Sigma^{\lambda}_{~\mu\lambda}\right)\Sigma^{\rho}_{~\nu\rho}
	\nonumber \\
	\!&+&\!
	\left(-B_{(2)}-I_{(2)}-2J_{(2)}\right)\eta^{\lambda\rho}\left(\partial_{\mu}\partial^{\nu}\Sigma^{\mu}_{~\lambda\rho}\right)\Sigma^{\alpha}_{~\nu\alpha}
	\nonumber \\
	\!&+&\!
	\left(-2F_{(2)}-I_{(2)}-2K_{(2)}\right)\eta^{\alpha\beta}\left(\partial_{\mu}\partial^{\nu}\Sigma^{\mu}_{~\lambda\alpha}\right)\Sigma^{\lambda}_{~\nu\beta}.
\end{eqnarray}
Here, a series of $A, B, C$ with subscripts, such as $A_{(1)}$, is a set of parameters. The action (\ref{the most general second-order perturbation action}) can give us the most general forms of $\mathcal{F}_{\mu\nu}$ and $\mathcal{G}_{\lambda}^{~\mu\nu}$.

\section{polarization modes of gravitational waves and gauge invariants}
\label{sec: 3}

To study the polarization modes of gravitational waves, we need to solve the linearized field equations of the theory. Before doing so, we first introduce the basic concepts of gauge-invariant perturbations and gravitational wave polarization modes in this section.

The gauge-invariant perturbation method can significantly simplify the linearized field equations. This approach allows us to reformulate the variables in the linearized field equations in terms of a set of gauge-invariant tensors, vectors, and scalars (in the sense of spatial rotations). Furthermore, by decoupling the linearized field equations, we obtain a series of tensor, vector, and scalar equations, each of which depends solely on the corresponding gauge-invariant tensor, vector, or scalar. This method has been elaborated in detail in Refs. \cite{Yu-Qi Dong,Y.Dong3,Y.Dong4,James M. Bardeen,Eanna E Flanagan,Y.Dong6}; here, we only summarize the necessary results.

It is known that perturbations can be uniquely decomposed as \cite{Yu-Qi Dong,Y.Dong6,James M. Bardeen,Eanna E Flanagan,Weinberg,Evans}
\begin{eqnarray}
	\label{decompose perturbations}
	h_{00}&=&h_{00}, \nonumber\\
	h_{0i}&=&\partial_{i}\gamma+\beta_{i},\nonumber\\
	h_{ij}&=&h^{TT}_{ij}+\partial_{i}\epsilon_{j}+\partial_{j}\epsilon_{i}
	+\frac{1}{3}\delta_{ij}H+(\partial_{i}\partial_{j}-\frac{1}{3}\delta_{ij}\Delta)\zeta,
	\nonumber\\
	\Sigma^{0}_{~00}&=&\Sigma^{0}_{~00}, \nonumber \\
	\Sigma^{i}_{~00}&=&\partial_{i}F+G_{i},\nonumber \\
	\Sigma^{0}_{~0i}&=&\partial_{i}M+N_{i},\nonumber \\
	\Sigma^{0}_{~ij}&=&S_{ij}+\partial_{i}C_{j}+\partial_{j}C_{i}+\partial_{i}\partial_{j}B+\delta_{ij}A,\nonumber \\
	\Sigma^{i}_{~0j}&=&U_{ij}+V_{ij}+\partial_{i}\bar{C}_{j}+\partial_{j}\bar{D}_{i}+\partial_{i}\partial_{j}\bar{B}+\delta_{ij}\bar{A},\nonumber \\
	\Sigma^{i}_{~jk}&=&B^{i}_{~jk}+\partial_{i}C_{jk}+\partial_{j}D_{ik}+\partial_{k}D_{ij}+\partial_{j}E_{ik}+\partial_{k}E_{ij}\nonumber \\
	&+&\partial_{i}\partial_{j}f_{k}+\partial_{i}\partial_{k}f_{j}+\partial_{j}\partial_{k}g_{i}+\delta_{jk}h_{i}+\delta_{ij}q_{k}+\delta_{ik}q_{j}\nonumber\\
	&+&\partial_{i}\partial_{j}\partial_{k}l+\delta_{jk}\partial_{i}m+\delta_{ij}\partial_{k}n+\delta_{ik}\partial_{j}n,
\end{eqnarray}
where some constraints have been imposed: 
\begin{eqnarray}
	&\partial_{i}\beta^{i}=\partial_{i}\epsilon^{i}=\partial_{i}G^i=\partial_{i}N^i=\partial_{i}C^i=\partial_{i}{\bar{C}}^i=\partial_{i}{\bar{D}}^i=\partial_{i}f^i=\partial_{i}g^i=\partial_{i}h^i=\partial_{i}q^i=0,\nonumber
	\\
	&h^{TTi}_{~~~~i}=S^{i}_{~i}=U^{i}_{~i}=V^{i}_{~i}=C^{i}_{~i}=D^{i}_{~i}=E^{i}_{~i}=0,\nonumber
	\\
	&\partial_{i}h^{TTi}_{~~~~j}=\partial_{i}S^{i}_{~j}=\partial_{i}U^{i}_{~j}=\partial_{i}V^{i}_{~j}=\partial_{i}C^{i}_{~j}=\partial_{i}D^{i}_{~j}=\partial_{i}E^{i}_{~j}=0,\nonumber
	\\
	&h^{TT}_{ij}=h^{TT}_{ji},~S_{ij}=S_{ji},~ U_{ij}=U_{ji},~ C_{ij}=C_{ji},\nonumber
	\\
	&D_{ij}=D_{ji},~ V_{ij}=-V_{ji},~ E_{ij}=-E_{ji},\nonumber
	\\
	&B^{i}_{~jk}=	B^{i}_{~kj},~ B^{i}_{~ik}=B^{i}_{~jk}\delta^{jk}=0,~ \partial_{i}B^{i}_{~jk}=\partial^{j}B^{i}_{~jk}=0.
\end{eqnarray}
Here, spatial indices are raised and lowered by $\delta_{ij}$, so there is no need to distinguish between upper and lower spatial indices. Considering that under a gauge transformation (infinitesimal coordinate transformation) $x^\mu \rightarrow x^\mu+\xi^{\mu}(x)$, we have 
\begin{eqnarray}
	\label{gauge transformation}
	h_{\mu\nu} \!\rightarrow\! h_{\mu\nu}-\partial_{\mu}\xi_{\nu}-\partial_{\nu}\xi_{\mu}, \quad
	\Sigma^{\lambda}_{~\mu\nu} \!\rightarrow\!
	\Sigma^{\lambda}_{~\mu\nu}+\partial_{\mu}\partial_{\nu}\xi^{\lambda}.
\end{eqnarray}
Then we can construct the following gauge-invariant perturbations \cite{Yu-Qi Dong}:
\begin{itemize}
	\item Transverse traceless third-order spatial tensor:
	\begin{eqnarray}
		\label{gauge invariant third-order spatial tensors}
		B^{i}_{~jk}.
	\end{eqnarray}
	
	\item Transverse traceless second-order spatial tensors:
	\begin{eqnarray}
		\label{gauge invariant second-order spatial tensors}
		h^{TT}_{ij},~S_{ij},~ U_{ij},~V_{ij},~ C_{ij},~D_{ij},~ E_{ij}.
	\end{eqnarray}
	
	\item Transverse spatial vectors:
	\begin{eqnarray}
		\label{gauge invariant spatial vectors}
		&N_{i},~C_{i},~\bar{C}_{i},~f_{i},~h_{i},~q_{i},
		\nonumber\\
		&\Xi_{i}\coloneqq \beta_{i}-\partial_{0}\epsilon_{i},
		\nonumber\\
		&\Omega_{i}\coloneqq G_{i}-\partial_{0}\bar{D}_{i},~
		\bar{\Omega}_{i}\coloneqq \bar{D}_{i}-\partial_{0}g_{i},~
		K_{i}\coloneqq\beta_{i}+\bar{D}_{i}.
	\end{eqnarray}
	
	\item Spatial scalars:
	\begin{eqnarray}
		\label{gauge invariant spatial scalars}
		&A,~\bar{A},~m,~n,
		\nonumber\\
		&\Theta\coloneqq \frac{1}{3}\left(H-\Delta\zeta\right),~
		\phi\coloneqq -\frac{1}{2}h_{00}+\partial_{0}\gamma-\frac{1}{2}\partial_{0}^{2}\zeta,
		\nonumber\\
		&\Psi\coloneqq \Sigma^{0}_{~00}-\partial_{0}M,~
		\bar{\Psi}\coloneqq M-\partial_{0}B,
		\nonumber\\
		&\Pi\coloneqq F-\partial_{0}\bar{B},~
		\bar{\Pi}\coloneqq \bar{B}-\partial_{0}{l},
		\nonumber\\
		&K \coloneqq -h_{00}+2M,~
		L \coloneqq H+2\Delta l.
	\end{eqnarray}
\end{itemize}

\begin{figure*}[htbp]
	\makebox[\textwidth][c]{\includegraphics[width=1.2\textwidth]{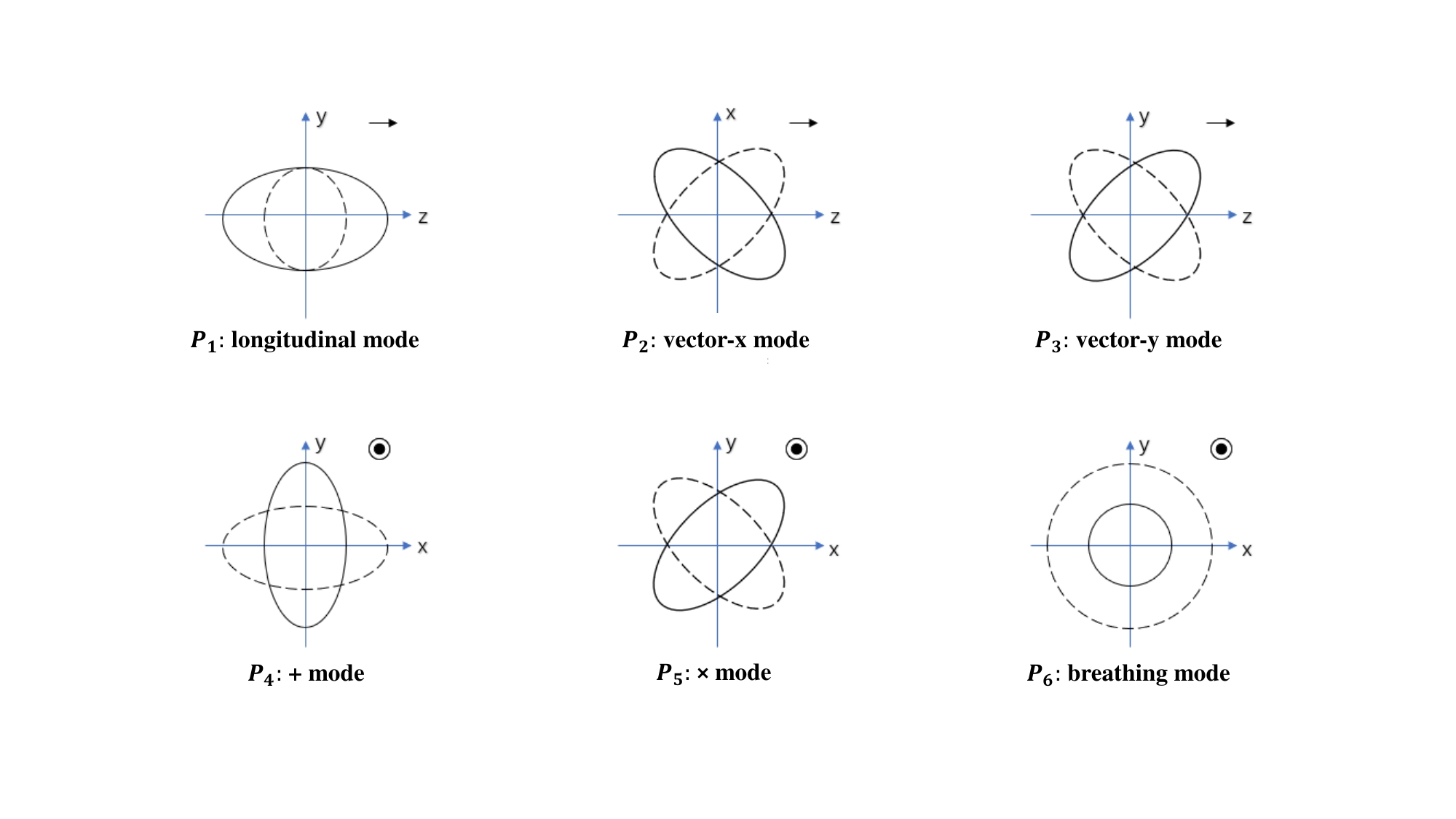}}
	\caption{The six polarization modes of gravitational waves \cite{Eardley}. Here, the gravitational wave propagates in the $+z$ direction. We have labeled the names of the six polarization modes in the figure. Based on the relationship between different polarization modes and gauge-invariant perturbations, the $+$ and $\times$ modes are referred to as tensor modes, the vector-$x$ and vector-$y$ modes are referred to as vector modes, and the breathing and longitudinal modes are referred to as scalar modes.}
	\label{fig: 1}
\end{figure*}

This concludes all the necessary knowledge about gauge-invariant perturbations. Now, we turn to the fundamental concepts of gravitational wave polarization modes. The polarization modes of gravitational waves are measured by the relative motion of test particles. For the case without hypermomentum, the equation of relative motion between two test particles is the well-known one:
\begin{eqnarray}
	\label{geodetic deviation equation}
	\frac{d^2 \eta^{i}}{dt^2}
	=
	-\widehat{R}^{i}_{~0j0}\eta^{j}.
\end{eqnarray}
It is always possible, without loss of generality, to consider the case where gravitational waves propagate along the $+z$ direction. Therefore, all gravitational waves discussed in this paper are set to propagate in the $+z$ direction. Under this assumption, the monochromatic plane wave leads to
\begin{eqnarray}
	\label{Ri0j0=AEeikx}
	\widehat{R}_{i0j0}=A_{0} E_{ij} e^{ikx}.
\end{eqnarray}
Here, $k^{\mu}$ is the four-wavevector, $A_{0}$ represents the wave intensity, and $E_{ij}$ is a $3 \times 3$ matrix encapsulating all polarization information of this plane wave, which satisfies
\begin{eqnarray}
	\label{EE=1}
	E_{ij}E^{ij}=1.
\end{eqnarray}
Due to the symmetry of the $(i, j)$ indices in $\widehat{R}_{i0j0}$, there can be at most six independent gravitational wave polarization modes in this case:
\begin{eqnarray}
	\label{P1-P6}
	\widehat{R}_{i0j0}=\begin{pmatrix}
		P_{4}+P_{6} & P_{5} & P_{2}\\
		P_{5}       & -P_{4}+P_{6}  & P_{3}\\
		P_{2}       &  P_{3}   &   P_{1}
	\end{pmatrix}.
\end{eqnarray}
The six modes are related to gauge-invariant perturbations as follows:
\begin{eqnarray}
	\label{P1-P6 gauge invariant}
	\begin{array}{l}
		P_{1}=\partial_{3}\partial_{3}\phi-\frac{1}{2}\partial_{0}\partial_{0}\Theta, \quad
		P_{2}=\frac{1}{2}\partial_{0}\partial_{3}\Xi_{1},\\
		P_{3}=\frac{1}{2}\partial_{0}\partial_{3}\Xi_{2},  \quad \quad\quad\quad\,\,
		P_{4}=-\frac{1}{2}\partial_{0}\partial_{0}h^{TT}_{11}, \\
		P_{5}=-\frac{1}{2}\partial_{0}\partial_{0}h^{TT}_{12}, \quad\quad~~~
		P_{6}=-\frac{1}{2}\partial_{0}\partial_{0}\Theta.
	\end{array}
\end{eqnarray}
Finally, the relative motion of test particles in these six modes is shown in Fig. \ref{fig: 1}. Figure \ref{fig: 1} also labels the names corresponding to these six modes.

\begin{figure*}[htbp]
	\makebox[\textwidth][c]{\includegraphics[width=0.7\textwidth]{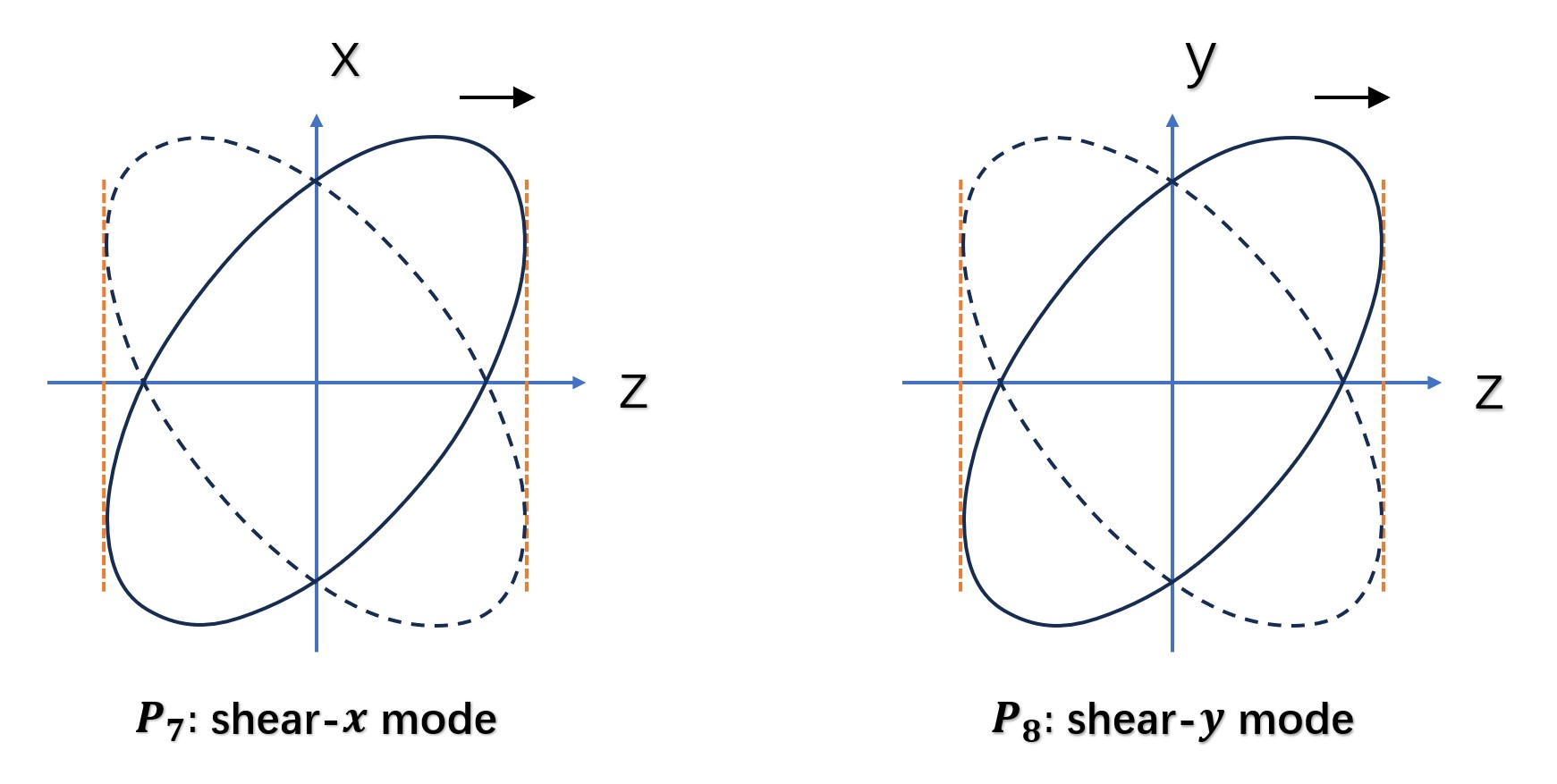}}
	\caption{The two new shear modes \cite{Yu-Qi Dong}.}
	\label{fig: 2}
\end{figure*}

For the case with hypermomentum, the equation of relative motion for the test particles is \cite{Yu-Qi Dong}
\begin{eqnarray}
	\label{equation of motion for etai}
	\frac{d^2 \eta^{i}}{dt^2}
	=
	-A^{i}_{~j}\eta^{j}
	\coloneqq
	-\widehat{R}^{i}_{~0j0}\eta^{j}-\partial_{j}N^{i}_{~00}\eta^{j}.
\end{eqnarray}
Here, 
\begin{eqnarray}
	\label{N lambda munu}
	N^{\lambda}_{~\mu\nu}\coloneqq \Gamma^{\lambda}_{~\mu\nu}-\widehat{\Gamma}^{\lambda}_{~\mu\nu}.
\end{eqnarray}
In this case, $A_{ij}$ is no longer symmetric with respect to the $(i, j)$ indices, which leads to two new polarization modes of gravitational waves:
\begin{eqnarray}
	\label{P1-P8}
	A_{ij}=\begin{pmatrix}
		P_{4}+P_{6} & P_{5} & P_{2}+P_{7}\\
		P_{5}       & -P_{4}+P_{6}  & P_{3}+P_{8}\\
		P_{2}       &  P_{3}   &   P_{1}
	\end{pmatrix}.
\end{eqnarray}
The two new polarization modes, $P_{7}$ and $P_{8}$, are referred to as shear modes \cite{Yu-Qi Dong}. The relative motion of test particles in these two modes is illustrated in Fig. \ref{fig: 2}.
The relationship between the polarization modes and gauge-invariant perturbations in this case is given by 
\begin{eqnarray}
	\label{P1-P8 gauge invariant}
	&P_{1}=\partial_{0}\partial_{0}\Theta
	-\partial_{3}\partial_{3}\Pi-\partial_{0}\partial_{3}\partial_{3}\bar{\Pi}
	-\frac{1}{2}\partial_{0}\partial_{0}L,~
	P_{6}=-\frac{1}{2}\partial_{0}\partial_{0}\Theta, \nonumber \\
	&P_{2}=\frac{1}{2}\partial_{0}\partial_{3}\Xi_{1},\quad P_{3}=\frac{1}{2}\partial_{0}\partial_{3}\Xi_{2},  \nonumber \\
	&P_{4}=-\frac{1}{2}\partial_{0}\partial_{0}h^{TT}_{11}, \quad
	P_{5}=-\frac{1}{2}\partial_{0}\partial_{0}h^{TT}_{12}, \nonumber \\
	&P_{7}=\partial_{3}\Omega_{1}
	-\partial_{0}\partial_{3}K_{1},\quad
	P_{8}=\partial_{3}\Omega_{2}
	-\partial_{0}\partial_{3}K_{2}.
\end{eqnarray}

\section{Gravitational wave polarization modes in the most general second-order symmetric teleparallel gravity}
\label{sec: 4}

In this section, we analyze the gravitational wave polarization modes in general symmetric teleparallel gravity, which can yields second-order field equations. The equations we need to solve are Eqs. (\ref{linear perturbation equations metric new})-(\ref{linear perturbation equations R new}), where the explicit expressions for $\mathcal{F}_{\mu\nu}$ and $\mathcal{G}_{\lambda}^{~\mu\nu}$ are obtained from the variation of the action (\ref{the most general second-order perturbation action}).

We first examine Eq. (\ref{linear perturbation equations R new}), which requires the curvature to be zero. When rewriting this equation in terms of the gauge-invariant perturbations from the previous section, we obtain 
\begin{eqnarray}
	\label{gauge-invariant perturbations when R=0}
   B^{i}_{~jk}\!&=&\!0,~S_{ij}=U_{ij}=V_{ij}=C_{ij}=D_{ij}=E_{ij}=0,
   \nonumber \\
   N_{i}\!&=&\!C_{i}=\bar{C}_{i}=\Omega_{i}=\bar{\Omega}_{i}=f_{i}=h_{i}=q_{i}=0,
   \nonumber \\
   \Psi\!&=&\!\bar{\Psi}=A=\bar{A}=\Pi=\bar{\Pi}=m=n=0.
\end{eqnarray}
It can be seen that for gauge-invariant tensor perturbations, only $h^{TT}_{ij}$ may propagate. Similarly, for vector perturbations, only $\Xi_{i}$ and $K_{i}$ may propagate, and for scalar perturbations, only $\Theta$, $\Phi$, $K$, and $L$ may propagate. Equation (\ref{gauge-invariant perturbations when R=0}) requires that all gauge-invariant perturbations composed solely of connection perturbations vanish, which is consistent with the fact that in symmetric teleparallel gravity, one can always choose the coincident gauge: $\Gamma^{\lambda}_{~\mu\nu}=0$ \cite{Lavinia Heisenberg3}.

Now, we consider the remaining Eqs. (\ref{linear perturbation equations metric new}) and (\ref{linear perturbation equations connection new}) by separately analyzing the tensor, vector, and scalar equations.

\subsection{Tensor modes}

The equation for the tensor perturbation is given by
\begin{eqnarray}
	\label{tensor equation}
	A_{(2)}\Box h^{TT}_{ij}=0.
\end{eqnarray}
Combining Eqs. (\ref{P1-P6 gauge invariant}) and (\ref{P1-P8 gauge invariant}), this implies that symmetric teleparallel gravity necessarily requires $A_{(2)} \neq 0$, and there must exist $+$ and $\times$ modes propagating at the speed of light, regardless of whether the test particles carry hypermomentum charge.

\subsection{Vector modes}

There are three vector equations:
\begin{eqnarray}
	\label{vector equation 1}
	4A_{(2)}\Delta\Xi_{i}
	+\left(C_{(1)}+E_{(1)}\right)
	\left[\partial_{0}^{2}K_{i}
	+\Delta\left(\Xi_{i}-K_{i}\right)\right]
	\!&=&\!0,
	\\
	\label{vector equation 2}
    4A_{(2)}\partial_{0}^{2}\Xi_{i}
    +\left(C_{(1)}+E_{(1)}\right)
    \left[
    \partial_{0}^{2}K_{i}
    +\Delta\left(\Xi_{i}-K_{i}\right)
    \right]
    \!&=&\!0,
    \\
    \label{vector equation 3}
    \left(C_{(1)}+E_{(1)}\right)\Box
    \left[\partial_{0}^{2}K_{i}
    +\Delta\left(\Xi_{i}-K_{i}\right)\right]
    \!&=&\!0.
\end{eqnarray}
Since these equations are linear, they can be transformed into an algebraic system by rewriting them in momentum space form to facilitate solving \cite{Yu-Qi Dong,Y.Dong5}. The results show that, 
when $4A_{(2)}+C_{(1)}+E_{(1)}=0$, $\Xi_{i}$ propagates at the speed of light. According to Eqs. (\ref{P1-P6 gauge invariant}) and (\ref{P1-P8 gauge invariant}), this implies the existence of vector-$x$ and vector-$y$ modes that propagate at the speed of light. For any other parameters,  $\Xi_{i}=0$ and vector-$x$ and vector-$y$ modes do not exist. In particular, when the test particles do not carry hypermomentum charge, vector mode gravitational waves emerge solely within a specific parameter space, as dictated by Eq. (\ref{P1-P6 gauge invariant}).

We find that when $C_{(1)}+E_{(1)} \neq 0$, there exists a solution for the perturbation $K_{i}$ and it propagates at the speed of light. However, when $C_{(1)}+E_{(1)}=0$, $K_{i}$ is not constrained by any equation. For the case where test particles do not carry hypermomentum, the condition $C_{(1)}+E_{(1)}=0$ does not pose any issues within the range we have examined, because, according to Eq. (\ref{P1-P6 gauge invariant}), the unconstrained $K_{i}$ is not observable. However, according to Eq. (\ref{P1-P8 gauge invariant}), we find that $K_{i}$ contributes to shear modes. Therefore, if a symmetric teleparallel gravity allows the existence of hypermomentum, it must satisfy $C_{(1)}+E_{(1)} \neq 0$, otherwise, observable pathological gravitational wave polarization modes will exist. In addition, for the case of test particles carrying hypermomentum, shear modes that propagating at the speed of light always exist.

\subsection{Scalar modes}
There are six scalar equations, which are relatively complex, so we have placed the detailed forms of these equations in Appendix \ref{app: A}. The properties of these scalar polarization modes depend on the parameter space, and their general analysis is very complex and tedious. However, two universal conclusions can still be extracted. The first one is that there always exists the following plane wave solution propagating at the speed of light: $L+K=0, \phi=\Theta=0$. Therefore, according to Eq. (\ref{P1-P8 gauge invariant}), in the case where the test particles have hypermomentum, there always exists a longitudinal mode propagating at the speed of light. The second one is that, owing to the homogeneity of the field equations and their formal Lorentz symmetry, all possible scalar modes must propagate at the speed of light.

\section{Examples of $f(Q)$ gravity and quadratic non-metricity gravity}
\label{sec: 5}

In this section, we study the gravitational wave polarization modes in two typical symmetric teleparallel gravity theories: $f(Q)$ gravity and quadratic non-metricity gravity.

\subsection{$f(Q)$ gravity}
The action of $f(Q)$ gravity is given by 
\begin{eqnarray}
	\label{fQ action}
	S=\int d^{4}x \sqrt{-g} f(Q),
\end{eqnarray}
where 
\begin{eqnarray}
	\label{Q}
	Q \coloneqq
	\frac{1}{4}Q_{\alpha\mu\nu}Q^{\alpha\mu\nu}
	-\frac{1}{2}Q_{\alpha\mu\nu}Q^{\mu\alpha\nu}
	-\frac{1}{4}Q_{\alpha\lambda}^{~~\lambda}Q^{\alpha\rho}_{~~\rho}
	+\frac{1}{2}Q_{\alpha\lambda}^{~~\lambda}Q^{\rho~\alpha}_{~\rho}.
\end{eqnarray}
Here, $f(Q)$ is an analytic function dependent on $Q$, and it can be expanded as 
\begin{eqnarray}
	\label{fQ=Q+Q2}
	f(Q)=\alpha Q+\beta Q^{2}+....
\end{eqnarray}
It can be seen that the parameters corresponding to $f(Q)$ theory are 
\begin{eqnarray}
	\label{fQ parameter}
	A_{(1)}=-\alpha,~
	B_{(1)}=\frac{1}{2}\alpha,~
	D_{(1)}=-\frac{1}{2}\alpha,~
	A_{(2)}=-\frac{1}{4}\alpha,
\end{eqnarray}
with all other parameters being zero.

By solving the linear perturbation equations, we find that this requires 
\begin{eqnarray}
	\label{fQ mode}
	\Xi_{i}=\Theta=\phi=0.
\end{eqnarray}
Therefore, in the case where test particles do not have hypermomentum, $f(Q)$ theory only has tensor modes ($+$ and $\times$ modes) propagating at the speed of light, without additional vector or scalar modes, which is consistent with the conclusion obtained in Ref. \cite{S Capozziello}. In the case with hypermomentum charge, according to Eq. (\ref{P1-P8 gauge invariant}), we find that, in addition to the tensor modes propagating at the speed of light, the theory always has unconstrained shear modes and the unconstrained longitudinal mode. This is physically unreasonable, and therefore, a viable $f(Q)$ theory must a priori assume that the matter field action is independent of the connection. 

However, assuming the absence of hypermomentum, Refs. \cite{D. Aguiar Gomes,Lavinia Heisenberg6} studied cosmological perturbations in $f(Q)$ theory with the typical scalar matter field $S_m=\int d^{4}x \left[-(\partial \Phi)^{2}-V(\Phi)\right]$ and found that ghost and strong coupling problems are widespread in $f(Q)$ gravity. It was suggested in Ref. \cite{Lavinia Heisenberg6} that introducing couplings between the connection and matter fields could be a potential solution to avoid these problems. However, our study demonstrates that this approach leads to observable pathological polarization modes of gravitational waves, effectively ruling out this possibility. In conclusion, whether or not hypermomentum is assumed, pathological issues typically arise in $f(Q)$ theory. Our study further highlights the importance of exercising caution when working with this theory.

\subsection{Quadratic non-metricity gravity}
Another typical theory is the quadratic non-metricity gravity, whose action includes all possible quadratic terms of $Q_{\lambda\mu\nu}$:
\begin{eqnarray}
	\label{quadratic non-metricity gravity action}
	S=\int d^{4}x \sqrt{-g}~\mathcal{L},
\end{eqnarray}
where,
\begin{eqnarray}	
	\mathcal{L}=
	c_{1}Q_{\alpha\mu\nu}Q^{\alpha\mu\nu}
	+c_{2}Q_{\mu\alpha\nu}Q^{\alpha\mu\nu}
	+c_{3}Q_{\mu\lambda}^{~~\lambda}Q^{\mu\rho}_{~~\rho}
	+c_{4}Q^{\lambda}_{~\lambda\mu}Q_{\rho}^{~\rho\mu}
	+c_{5}Q_{\mu\lambda}^{~~\lambda}Q_{\rho}^{~\rho\mu}.
\end{eqnarray}
References \cite{Ismail Soudi,Manuel Hohmann} have already conducted a certain level of investigation into the gravitational wave polarization modes of this theory in the absence of hypermomentum. The parameters corresponding to this theory are as follows:
\begin{eqnarray}
	\label{quadratic non-metricity parameter}
	A_{(1)}=2c_{2},~
	B_{(1)}=c_{5},~
	C_{(1)}=4c_{1}+2c_{2},~
	D_{(1)}=4c_{3}+c_{5},~
	E_{(1)}=2c_{4},~
	A_{(2)}=-c_{1},
\end{eqnarray}
and all other parameters are zero. In this case, scalar polarization modes of gravitational waves still depend on the parameter space.

In the literature on quadratic non-metricity gravity, the following two conditions for these parameters are frequently mentioned \cite{Lavinia Heisenberg2,Lavinia Heisenberg3,Jose Beltran Jimenez}:\\
carameter (a)
\begin{eqnarray}
	\label{quadratic non-metricity parameter a}
	c_2+c_4=-2c_1,~
	c_3=-c_1,~
	c_5=2c_1;
\end{eqnarray}
carameter (b) 
\begin{eqnarray}
	\label{quadratic non-metricity parameter b}
	c_2+c_4=-2c_1,~
	c_3=-\frac{3}{8}c_1,~
	c_5=2c_1.
\end{eqnarray}
By imposing the coincident gauge $\Sigma^{\lambda}_{~\mu\nu}=0$ at the level of the second-order perturbation action while retaining only the variable $h_{\mu\nu}$, the above two  conditions respectively endow the second-order perturbation action with the gauge symmetry and the Weyl Transverse Diffeomorphism (WTDiff) symmetry \cite{Lavinia Heisenberg2,Lavinia Heisenberg3,Jose Beltran Jimenez} in form. We now analyze the polarization modes of gravitational waves under these two conditions. 

For condition (a), we have $C_{(1)}+E_{(1)}=4c_1+2c_2+2c_4=0$. Therefore, there must be an unconstrained $K_{i}$ in the theory, leading to a pathological situation with hypermomentum. For the case without hypermomentum, there are no vector-mode gravitational waves. For scalar modes, it gives the solution $\phi=\Theta=0$, while imposing no constraints on $K$ and $L$. Therefore, in the absence of hypermomentum, scalar-mode gravitational waves do not exist. In conclusion, the theory only includes $+$ and $\times$ modes that propagate at the speed of light. For condition (b), the same reasoning leads to the conclusion that only the case with no hypermomentum is reasonable. Similarly, in the absence of hypermomentum charge, there are no vector-mode gravitational waves. In this case, there is no scalar mode, although $L$ and $K$ can propagate. In this condition, this theory predicts only $+$ and $\times$ modes propagating at the speed of light.

Regarding quadratic non-metricity gravity, one additional point needs to be emphasized. It is evident that the linear field equations of general symmetric teleparallel gravity involve six free parameters—$A_{(1)},B_{(1)}, C_{(1)}, D_{(1)}, E_{(1)}$, and $A_{(2)}$—whereas quadratic non-metricity gravity contains only five independent parameters. This indicates that even at the linear order, there are theories that can not be encompassed by quadratic non-metricity gravity, which includes all possible quadratic terms of $Q_{\lambda\mu\nu}$. However, we note that the following action can encompass all six free parameters:
\begin{eqnarray}
	\label{quadratic non-metricity gravity action new}
	S=\int d^{4}x \sqrt{-g}~\mathcal{L},
\end{eqnarray}
where,
\begin{eqnarray}	
	\mathcal{L}=
	c_{0} R
	+c_{1}Q_{\alpha\mu\nu}Q^{\alpha\mu\nu}
	+c_{2}Q_{\mu\alpha\nu}Q^{\alpha\mu\nu}
	+c_{3}Q_{\mu\lambda}^{~~\lambda}Q^{\mu\rho}_{~~\rho}
	+c_{4}Q^{\lambda}_{~\lambda\mu}Q_{\rho}^{~\rho\mu}
	+c_{5}Q_{\mu\lambda}^{~~\lambda}Q_{\rho}^{~\rho\mu}.
\end{eqnarray}
Compared to quadratic non-metricity gravity, it has an additional scalar curvature term. The parameters corresponding to this theory are as follows:
\begin{eqnarray}
	\label{quadratic non-metricity parameter new}
	A_{(1)}\!&=&\!-c_{0}+2c_{2},~
	B_{(1)}=\frac{1}{2}c_{0}+c_{5},~
	C_{(1)}=4c_{1}+2c_{2},
	\nonumber \\
	D_{(1)}\!&=&\!-\frac{1}{2}c_{0}+4c_{3}+c_{5},~
	E_{(1)}=2c_{4},~
	A_{(2)}=-c_{1}.
\end{eqnarray}
A view holds that symmetry teleparallel gravity requires the curvature to be zero, so $R$ does not contribute to the action. However, it is important to emphasize that the condition of zero curvature is imposed at the level of the field equations, not at the level of the action. Therefore, one cannot naively impose constraints on the action, but rather must carefully consider and justify them; otherwise, errors may arise.

\section{Conclusion}
\label{sec: 6}

In this paper, we studied the polarization modes of gravitational waves in the most general symmetric teleparallel gravity, where second-order field equations can be derived, both with and without hypermomentum. We found that, in the case without hypermomentum, the theory always contains two tensor polarization modes of gravitational waves, the $+$ and $\times$ modes, propagating at the speed of light. Unless in an extremely special condition for the parameters: $4A_{(2)}+C_{(1)}+E_{(1)}=0$, the theory generally does not have vector modes. The existence of scalar modes depends on the choice of parameters, but the possible scalar gravitational waves can only propagate at the speed of light. In the case with hypermomentum, there are always two tensor modes propagating at the speed of light. For vector modes, shear modes must either propagate at the speed of light or be pathological (they are not constrained by the linearized field equations). The vector-$x$ and vector-$y$ modes are only permissible under the specific parameter constraint $4A_{(2)}+C_{(1)}+E_{(1)}=0$. While scalar modes similarly depend on system parameters, but there always exists a longitudinal mode that propagates at the speed of light regardless of parameter values. Our analysis reveals that, in the presence of hypermomentum, the additional shear modes and the longitudinal mode are universally present. This distinctive characteristic provides an important criterion for distinguishing symmetric teleparallel gravity from other gravitational theories.

We also specifically considered two typical symmetric teleparallel gravity theories: $f(Q)$ gravity and quadratic non-metricity gravity. For $f(Q)$ gravity, we found that assuming any coupling between the matter field and the connection, i.e., the presence of hypermomentum, renders the theory physically unreasonable. Therefore, within the framework of $f(Q)$ theory, the matter field must couple solely to the metric, and in this case, the theory has only two tensor modes that propagate at the speed of light. Combining the analysis of the ghost and strong coupling problems in the context of cosmological perturbations from Refs. \cite{D. Aguiar Gomes,Lavinia Heisenberg6}, our results further suggest that $f(Q)$ theory must be treated with caution, as it may exhibit pathological behavior.

For quadratic non-metricity gravity, we performed a detailed analysis of its gravitational wave polarization modes under two specific parameter conditions. We found that under both conditions, the assumption of the existence of hypermomentum can not be made. In the absence of hypermomentum, only tensor modes propagating at the speed of light exist. We also pointed out that, at the linear level, quadratic non-metricity gravity can not cover the most general linearized field equations. By introducing the scalar curvature $R$ into the action, we found a class of theories that can cover the most linearized field equations at the linear level. The nontrivial contribution of $R$ may lead to unexpected results.

\section*{Acknowledgments}
This work is supported in part by the National Key Research and Development Program of China (GrantNo. 2020YFC2201503), the National Natural Science Foundation of China (Grants No. 123B2074, No. 12475056,  and No. 12247101), Gansu Province's Top Leading Talent Support Plan, the Fundamental Research Funds for the Central Universities (Grant No. lzujbky-2024-jdzx06), the Natural Science Foundation of Gansu Province (No. 22JR5RA389), and the `111 Center' under Grant No. B20063.

\appendix
\section{The specific forms of scalar equations}
\label{app: A}

\begin{eqnarray}
	\label{scala equation 1}
	&\left(-\frac{1}{2}A_{(1)}-\frac{3}{2}B_{(1)}-\frac{1}{2}C_{(1)}-\frac{1}{2}D_{(1)}-E_{(1)}\right)\partial_{0}^{2}K
	\nonumber \\
	&+\left(\frac{1}{2}B_{(1)}+\frac{1}{2}C_{(1)}+\frac{1}{2}D_{(1)}+\frac{1}{2}E_{(1)}\right)\Delta K
	+\left(-A_{(1)}-2B_{(1)}-E_{(1)}\right)\Delta \phi
	\nonumber\\
	&+\left(-\frac{1}{2}A_{(1)}-\frac{3}{2}B_{(1)}-\frac{1}{2}D_{(1)}-\frac{1}{2}E_{(1)}\right)\partial_{0}^{2}L
	+\left(\frac{1}{2}B_{(1)}+\frac{1}{2}D_{(1)}\right)\Delta L
	\nonumber \\
	&+\left(\frac{3}{2}A_{(1)}+3B_{(1)}+\frac{3}{2}E_{(1)}\right)\partial_{0}^{2}\Theta
	+\left(4A_{(2)}+C_{(1)}-A_{(1)}-2B_{(1)}\right)\Delta \Theta=0,
\end{eqnarray}				
\begin{eqnarray}
	\label{scalar equation 2}
	&\left(-\frac{1}{2}A_{(1)}-\frac{1}{4}C_{(1)}-\frac{3}{4}E_{(1)}-B_{(1)}\right)\partial_{0}K
	+\left(\frac{1}{4}C_{(1)}+\frac{1}{4}E_{(1)}\right)\frac{\Delta}{\partial_{0}}K
	\nonumber \\
	&
	+\left(\frac{1}{2}C_{(1)}+\frac{1}{2}E_{(1)}\right)\partial_{0}\phi
	+\left(-\frac{1}{2}C_{(1)}-\frac{1}{2}E_{(1)}\right)\frac{\Delta}{\partial_{0}}\phi
	\nonumber \\
	&+\left(\frac{1}{4}C_{(1)}+\frac{1}{4}E_{(1)}\right)\frac{\partial_{0}^{3}}{\Delta}L
	+\left(-\frac{1}{2}A_{(1)}-\frac{1}{4}C_{(1)}-B_{(1)}-\frac{3}{4}E_{(1)}\right)\partial_{0}L
	\nonumber \\
	&-\left(\frac{3}{4}C_{(1)}+\frac{3}{4}E_{(1)}\right)\frac{\partial_{0}^{3}}{\Delta}\Theta
	+\left(4A_{(2)}+\frac{7}{4}C_{(1)}+\frac{7}{4}E_{(1)}\right)\partial_{0}\Theta=0,
\end{eqnarray}			
\begin{eqnarray}
	\label{scalar equation 3}
	&\left(-\frac{1}{2}A_{(1)}-B_{(1)}-\frac{1}{2}E_{(1)}\right)K
	\nonumber \\
	&
	+\left(\frac{1}{2}C_{(1)}+\frac{1}{2}E_{(1)}\right)\frac{\partial_{0}^{2}}{\Delta}L
	+\left(-\frac{1}{2}A_{(1)}-\frac{1}{2}C_{(1)}-E_{(1)}-B_{(1)}\right)L
	\nonumber \\
	&+\left(-\frac{3}{2}C_{(1)}-\frac{3}{2}E_{(1)}\right)\frac{\partial_{0}^{2}}{\Delta}\Theta
	+\left(2A_{(2)}+2C_{(1)}+2E_{(1)}\right)\Theta=0,
\end{eqnarray}				
\begin{eqnarray}
	\label{scalar equation 4}
	&\left(\frac{1}{2}B_{(1)}+\frac{1}{2}D_{(1)}\right)\partial_{0}^{2}K
	+\left(-\frac{1}{2}B_{(1)}-\frac{1}{2}D_{(1)}\right)\Delta K
	\nonumber \\
	&+\left(-4A_{(2)}-C_{(1)}+A_{(1)}+2B_{(1)}\right)\Delta\phi
	\nonumber \\
	&+\left(\frac{1}{2}B_{(1)}+\frac{1}{2}D_{(1)}\right)\partial_{0}^{2}L
	+\left(-\frac{1}{2}B_{(1)}-\frac{1}{2}D_{(1)}\right)\Delta L
	\nonumber \\
	&+\left(4A_{(2)}+\frac{3}{2}C_{(1)}-\frac{3}{2}A_{(1)}-3B_{(1)}\right)\partial_{0}^{2}\Theta
	+\left(-2A_{(2)}-C_{(1)}+A_{(1)}+2B_{(1)}\right)\Delta\Theta
	\nonumber\\
	&=0,
\end{eqnarray}	
\begin{eqnarray}
	\label{scalar equation 5}
	&\partial_{0}^{2}
	\left[
	\left(
	-3B_{(1)}-2E_{(1)}-D_{(1)}-C_{(1)}-A_{(1)}
	\right)\partial_{0}K
	+\left(B_{(1)}+E_{(1)}\right)\frac{\Delta}{\partial_{0}}K
	\right.
	\nonumber \\
	&\left.
	-\left(2B_{(1)}+2E_{(1)}\right)\frac{\Delta}{\partial_{0}}\phi
	+\left(3B_{(1)}+3E_{(1)}\right)\partial_{0}\Theta
	+\left(-2B_{(1)}-E_{(1)}-D_{(1)}\right)\partial_{0}L
	\right]
	\nonumber \\
	&+2\partial_{0}\Delta\left[
	\left(
	\frac{B_{(1)}}{2}+\frac{D_{(1)}}{2}+\frac{E_{(1)}}{4}+\frac{3C_{(1)}}{4}+\frac{A_{(1)}}{2}
	\right)K
	+\left(-B_{(1)}-\frac{E_{(1)}}{2}-\frac{C_{(1)}}{2}-A_{(1)}\right)\phi
	\right.
	\nonumber \\
	&-\left(
	\frac{B_{(1)}}{2}+\frac{E_{(1)}}{4}+\frac{A_{(1)}}{2}+\frac{C_{(1)}}{4}
	\right)\frac{\partial_{0}^{2}}{\Delta}L
	+\left(
	B_{(1)}+\frac{D_{(1)}}{2}+\frac{E_{(1)}}{2}
	\right)L
	\nonumber \\
	&+\left.
	\left(
	\frac{3B_{(1)}}{2}+\frac{3E_{(1)}}{4}+\frac{3A_{(1)}}{2}+\frac{3C_{(1)}}{4}\right)\frac{\partial_{0}^{2}}{\Delta}\Theta
	-\left(2B_{(1)}+E_{(1)}\right)\Theta
	\right]
	\nonumber \\
	&+\Delta^{2}\left[
	-\frac{1}{2}C_{(1)}\frac{1}{\partial_{0}}K
	+C_{(1)}\frac{1}{\partial_{0}}\phi
	+\left(A_{(1)}+\frac{C_{(1)}}{2}\right)\frac{\partial_{0}}{\Delta}L
	+\left(-3A_{(1)}-\frac{3C_{(1)}}{2}\right)\frac{\partial_{0}}{\Delta}\Theta
	\right]
	\nonumber \\
	&\Delta\left[
	\left(B_{(1)}+E_{(1)}\right)\partial_{0}K
	-\frac{E_{(1)}}{2}\frac{\Delta}{\partial_{0}}K
	+E_{(1)}\frac{\Delta}{\partial_{0}}\phi
	+\left(B_{(1)}+\frac{E_{(1)}}{2}\right)\partial_{0}L
	+\left(A_{(1)}-\frac{3E_{(1)}}{2}\right)\partial_{0}\Theta
	\right]
	\nonumber \\
	&=0,
\end{eqnarray}	
\begin{eqnarray}
	\label{scalar equation 6}
	&\partial_{0}^{2}\left[
	\left(-A_{(1)}-B_{(1)}-\frac{C_{(1)}}{2}-\frac{E_{(1)}}{2}\right)K
	+\left(C_{(1)}+E_{(1)}\right)\phi
	+\frac{1}{2}\left(C_{(1)}+E_{(1)}\right)\frac{\partial_{0}^{2}}{\Delta}L
	\right.
	\nonumber \\
	&\left.
	-\left(B_{(1)}+E_{(1)}\right)L
	+2E_{(1)}\Theta
	-\frac{3}{2}\left(C_{(1)}+E_{(1)}\right)\frac{\partial_{0}^{2}}{\Delta}\Theta
	\right]
	\nonumber \\
	&+2\partial_{0}\left[
	\left(-B_{(1)}-\frac{E_{(1)}}{2}-\frac{D_{(1)}}{2}\right)\partial_{0}K
	+\left(\frac{B_{(1)}}{2}+\frac{E_{(1)}}{4}\right)\frac{\Delta}{\partial_{0}}K
	-\left(B_{(1)}+\frac{E_{(1)}}{2}\right)\frac{\Delta}{\partial_{0}}\phi
	\right.
	\nonumber \\
	&\left.
	-\left(\frac{B_{(1)}}{2}+\frac{D_{(1)}}{2}+\frac{E_{(1)}}{4}\right)\partial_{0}L
	+\left(\frac{3B_{(1)}}{2}+\frac{3E_{(1)}}{4}-\frac{C_{(1)}}{2}\right)\partial_{0}\Theta
	\right]
	\nonumber \\
	&+\Delta\left[
	\left({A_{{1}}}+\frac{C_{(1)}}{2}\right){K}
	-\left(2A_{(1)}+{C_{(1)}}\right){\phi}
	-\left({A_{(1)}}+\frac{3C_{(1)}}{2}\right)\frac{\partial_{0}^{2}}{\Delta}L
	+\left({3A_{(1)}}+\frac{9C_{(1)}}{2}\right)\frac{\partial_{0}^{2}}{\Delta}\Theta
	\right]
	\nonumber \\
	&+\Delta\left[
	\left(A_{(1)}+C_{(1)}\right){L}
	-3\left(A_{(1)}+C_{(1)}\right){\Theta}
	\right]
	\nonumber \\
	&+\Delta\left[
	\left(B_{(1)}+\frac{E_{(1)}}{2}\right)K
	-E_{(1)}\phi
	-\frac{E_{(1)}}{2}\frac{\partial_{0}^{2}}{\Delta}L
	+\left(B_{(1)}+E_{(1)}\right)L
	+\frac{3E_{(1)}}{2}\frac{\partial_{0}^{2}}{\Delta}\Theta
	+\left(A_{(1)}-2E_{(1)}\right)\Theta
	\right]
	\nonumber \\
	&+2\Delta\left[
	\left(\frac{B_{(1)}}{2}+\frac{D_{(1)}}{2}+\frac{E_{(1)}}{4}\right)K
	-\left(B_{(1)}+\frac{E_{(1)}}{2}\right)\phi
	-\left(\frac{B_{(1)}}{2}+\frac{E_{(1)}}{4}\right)\frac{\partial_{0}^{2}}{\Delta}L
	\right.
	\nonumber \\
	&\left.
	+\left(B_{(1)}+\frac{E_{(1)}}{2}+\frac{D_{(1)}}{2}\right)L
	+\left(\frac{3B_{(1)}}{2}+\frac{3E_{(1)}}{4}\right)\frac{\partial_{0}^{2}}{\Delta}\Theta
	+\left(-2B_{(1)}-E_{(1)}+\frac{C_{(1)}}{2}\right)\Theta
	\right]
	\nonumber \\
	&=0.
\end{eqnarray}

\end{document}